# Macroscopic wall pressure and microscopic contact load in crowds without egress: social-group cohesion and boundary buffering

Bo-Shiun Shen[1, 2] and Son-Hsien Chen [1, *]

[1] *Department of Applied Physics and Chemistry, University of Taipei, Taipei 100234, Taiwan*
[2] *Graduate Institute of Photonics and Optoelectronics, National Taiwan University, Taipei 10617, Taiwan*

Crowd safety in confined venues is commonly studied through evacuation efficiency or pre-collision avoidance, whereas the direct mechanical risks of crowds without egress, such as those arising during dense indoor gatherings or temporary lockdowns in concert halls and foyers, remain largely unexplored. In this paper, we sequentially examine an elastic reorientation model (ERM), social force model (SFM) interactions, and their combined dynamics. Post-collision cognitive modulation is introduced through social-group cohesion $\gamma_g$ and wall buffering $\gamma_w$, while mechanical safety is quantified using the macroscopic wall line pressure $P_{\text{wall}}$ and the microscopic maximum per-agent collision impulse $\delta p_{\max}$. Within the ERM, cohesion and wall buffering generally suppress $P_{\text{wall}}$ by retaining agents in the bulk, but large groups develop a pronounced high-$\delta p_{\max}$ hazard window at intermediate cohesion. Near complete cohesion, $\gamma_g \to 1$, local pairing suppresses the cluster growth and, together with the shift of cluster kinetic energy from relative to center-of-mass motion, reduces $\delta p_{\max}$. SFM pushing and sliding substantially amplify $\delta p_{\max}$, with further enhancement when agent-wall interactions coexist, whereas active driving increases $P_{\text{wall}}$ through near-wall accumulation. The combined dynamics produces a wall-pressure/contact-load trade-off ($P$-$p$ trade-off), reflecting the competing effects of the ERM and SFM on these two safety metrics. Finite-size scaling further reveals an independent-agent-induced phase boundary at $\gamma_w = 0.5$, characterized by a susceptibility discontinuity, and a grouped-agent-induced continuous phase boundary along a finite segment of $(1-\gamma_w)(1-\gamma_g) = 0.5$. The latter is characterized by a divergent susceptibility and terminates at a critical point. Both phase structures vanish in the social-force-free ERM, confirming that they emerge from the coupled ERM+SFM dynamics. The findings provide mechanistic guidance for group-configuration assignment, behavioral coordination, and event safety planning in high-density venues without egress.

## I. INTRODUCTION

Active particles differ from passive ones in that each unit continuously converts free energy into directed motion [1, 2], sustaining intrinsically nonequilibrium dynamics even in steady states [3]. As a result, active systems display collective phenomena absent in equilibrium matter, including motility-induced phase separation [4] and anomalous mechanical pressures that depend explicitly on boundary properties rather than solely on intrinsic state variables [5]. A crowd may be viewed as an assembly of social agents governed by coupled mechanical, psychological, and social interactions [6], and thus represents a particular realization of active matter, especially under strong geometric confinement. This perspective is useful for understanding and mitigating catastrophic crowd events, such as the Hajj stampedes [7], the Love Parade disaster [8], and the Itaewon crowd crush [9].

Crowd dynamics has been explored using a broad range of modeling approaches, including collective motion models [10], force-based individual models [6, 11–14], continuum descriptions [8, 15, 16], and discrete rule-based models such as cellular automata [17, 18]. These frameworks have been validated through controlled experiments [19–21], video-based analyses of real events [7, 22], and hybrid methods combining simulation with virtual reality [23–25], discrete-element mechanics [26], or data-driven approaches [27].

Crowd dynamics has also been studied in open systems [7, 8, 20, 22, 25, 28] and in quasi-open evacuation settings with limited egress [21, 29–34]. In open settings, increasing density induces transitions from laminar flow to stop-and-go waves [7, 35], and ultimately to turbulent motion [7, 8, 16, 22], characterized by intermittent, force-chain-mediated displacements [7, 20, 22]. Hazards in this regime are therefore commonly assessed using kinematic indicators such as density $\rho$, flux $\rho v$ [19, 36, 37], and "pressure" proxies based on speed variance, $\rho\text{Var}(v)$ [7].

By contrast, in quasi-open evacuation settings, safety is primarily evaluated through egress efficiency. Evacuation performance is shaped by bottleneck effects including bursty outflow [29, 35] and the faster-is-slower phenomenon [29], whereby increased competitiveness degrades flow due to congestion and clogging. Outcomes further depend on behavioral heterogeneity and group structure [21, 30, 33]; for example, mixing agents of different speeds can intensify congestion, while explicit group-maintenance or coordination rules may delay exit near bottlenecks [21, 30].

In addition to egress efficiency, pre-collision (or pre-contact) avoidance has been extensively studied using predictive motion-planning methods such as Velocity Obstacle (VO) and Reciprocal Velocity Obstacle (RVO) algorithms [38–40], as well as force-based models. Within the latter class, the social force model (SFM) introduced by Helbing and Molnár has become a standard framework for dense and panic-driven crowds [6, 29]. Numerous SFM variants have been developed to incorporate relative-velocity effects, anticipatory avoidance, and empirically calibrated interaction rules [13, 27, 31, 32, 41–46]. In particular, unphysical back-and-forth oscillations near congested exits can be mitigated by psychologically motivated predictors, such as predicted-position criteria [32] or time-headway rules for foreseen collisions [31]. Social-group effects are likewise essential for realistic crowd modeling; about 70% of urban traffic occurs in social groups [28], increasing to roughly 95% at large events [47]. However, strictly closed (no-egress) crowds remain understudied from a *direct* mechanical-safety perspective, particularly when cohesive social groups and *post-collision* cognitive modulation are included.

* sonhsien@utaipei.edu.tw

In this paper, we study a contact-rich regime of crowd dynamics in which a spatially confined crowd evolves without egress, motivated by temporary lockdown scenarios in indoor venues such as halls or foyers. In the dense regime considered here, collisions are unavoidable. We therefore incorporate two post-collision behavioral mechanisms: social-group cohesion and boundary buffering that reduces loading on walls. Mechanical safety is characterized using two direct contact-based measures: the stabilized macroscopic wall pressure $P_{\rm wall}$ and the microscopic extreme per-agent contact load $\delta p_{\rm max}$, defined as the maximum per-agent impulse. The respective roles of event-driven cognitive modulation and continuous force-based regulation are extracted by examining (*i*) reorientation-dominated dynamics, (*ii*) SFM-based interactions, and (*iii*) their combination. The coupled dynamics produces a wall-pressure/contact-load trade-off and distinct thermodynamic phase transitions in $P_{\rm wall}$, as established through finite-size scaling. The findings provide mechanistic principles for mitigating mechanical loading, supporting safer planning and management of activities in confined venues. Moreover, the proposed physical-cognitive framework with post-collision modulations offers a minimal and extensible route toward behaviorally informed theoretical crowd modeling.

The remainder of this paper is organized as follows. Section II describes the physical system and introduces the modeling framework. This includes the formulation of elastic, collision-triggered social-group cohesion in Sec. II A and inelastic, non-contact social repulsion based on the SFM in Sec. II B. Section III presents the main findings and demonstrates that the results are robust against varying initial conditions (ICs). The individual and combined effects of collision-triggered reorientation and continuous social-force interactions are then systematically analyzed in Secs. III A, III B, and III C, respectively. Finally, Sec. IV concludes this study.

## II. MODEL AND FORMALISM

As illustrated in the schematic (Fig. 1), the system consists of $N$ agents enclosed within a two-dimensional square domain of side length $L$. The boundary comprises four walls, denoted by $w \in \{l, r, b, t\}$ for the left, right, bottom, and top walls, respectively. Each agent is modeled as a rigid disk of radius $R$ moving in the $x$-$y$ plane. This rigid-body approximation neglects body deformation and the associated dissipative losses, as further discussed below in relation to the conservative interpretation of $\delta p_{\rm max}$ in Sec. II A. Our analysis focuses on two primary mechanical safety metrics: (*i*) the *macroscopic* (time-coarse-grained) wall line pressure,

$$P_{\rm wall} \equiv \frac{1}{4L} \sum_{w} \frac{\Delta p_w}{\Delta t}, \tag{1}$$

where $\Delta p_w$ is the momentum transferred to wall $w$ during a time interval $\Delta t \gg f_w^{-1}$ (where $f_w$ denotes the agent-wall collision frequency), and the sum extends over all four walls, and (*ii*) the *microscopic* per-agent momentum change $\delta p$ arising from each collision event, from which the extreme contact load $\delta p_{\rm max}$ is obtained as the maximum recorded value. By introducing a contact duration $\delta t_c$, an effective per-agent contact force can be estimated as $\delta p/\delta t_c$, where $\delta p$ is computed from the collision-induced velocity updates. These updates are governed by distinct behavioral mechanisms, including hard-core collisions, social-group cohesion, and continuous social-force driving and repulsion, as formulated below. An overview of these modeling layers is provided in Table I. To prevent the system from relaxing into a trivial, zero-pressure rest state due to purely dissipative interactions, we impose two distinct activity constraints. First, for the collision-triggered mechanisms (Sec. II A), we restrict the velocity updates to pure directional reorientations, thereby enforcing strict kinetic-energy conservation. Second, in the continuous SFM framework (Sec. II B), we incorporate an active driving term that maintains motion toward a finite desired velocity $\vec{v}^*$.

### A. Collision-triggered social-group reorientation model

Our post-collision updates are formulated as speed-preserving directional reorientation rules. This isolates collision-triggered direction selection from continuous speed-control mechanisms, avoiding the introduction of additional, context-dependent dissipation parameters. In the absence of active driving, any post-collision dissipation would only reduce subsequent momentum changes; consequently, for a given tolerance $p^*$, the simulated $\delta p_{\rm max}$ serves as a conservative estimate of the extreme load, and $\delta p_{\rm max} < p^*$ provides a sufficient mechanical-safety condition within this modeling framework.

We begin with the mechanisms involving agent-agent and agent-wall collisions. For inter-agent collisions, the velocity after each collision is updated according to standard scattering kinematics,

$$\vec{v}_{i,c}' = \vec{v}_i - \frac{1+e}{m_i} \mu_{ij} \left( \delta\vec{v}_{ij} \cdot \hat{n}_{ij} \right) \hat{n}_{ij}. \tag{2}$$

TABLE I. Classification of modeling layers used in this work. Collision-triggered mechanisms are event-driven, while social force model interactions are persistent. The former constrains the system via kinetic-energy conservation, while the latter constrains speeds via relaxation toward a desired speed.

| Modeling layer | Symbol (strength) | Kinetic energy conserved | Temporal nature | Spatial action |
|---|---|---|---|---|
| Standard elastic collision | Baseline | Yes | Instantaneous; triggered by agent-agent or agent-wall contact | Discrete directional update |
| Collision-triggered social-group cohesion | $\gamma_g$ | Yes | Instantaneous; triggered by agent-agent collisions | Discrete directional reorientation |
| Collision-triggered wall buffering | $\gamma_w$ | Yes | Instantaneous; triggered by agent-agent collisions near walls | Discrete directional reorientation |
| Agent-agent (agent-wall) social pushing force | $\alpha$ ($\alpha_w$) | No | Persistent | Continuous force field |
| Agent-agent (agent-wall) social sliding force | $\beta$ ($\beta_w$) | No | Persistent | Continuous force field |
| Social driving force | $F^d$ | No | Persistent | Continuous force field |

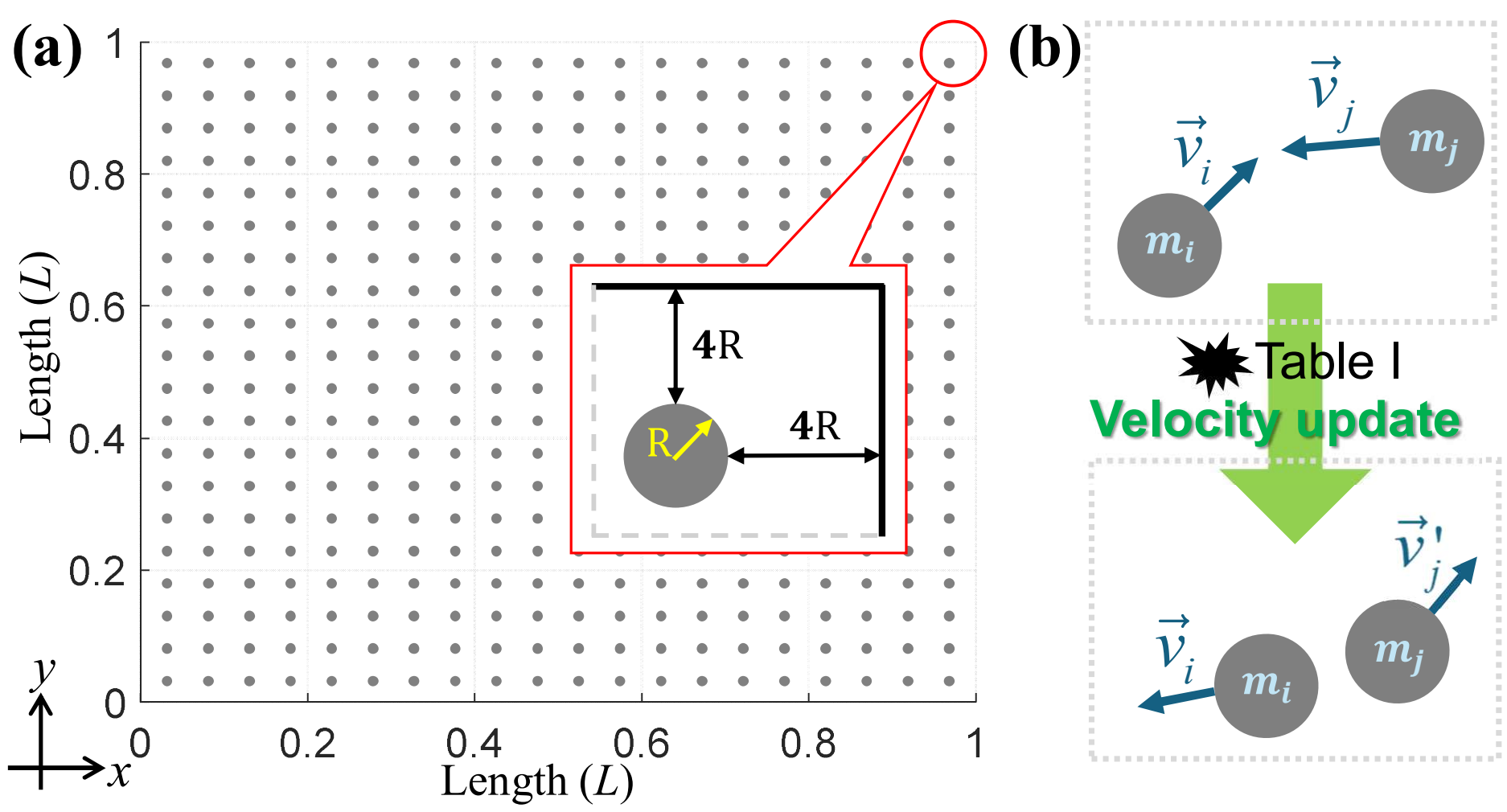

FIG. 1. (a) Schematic of the studied system and (b) illustration of the velocity updates. In (a), the bounded domain is of size $L \times L$ in the $x$-$y$ plane, and contains a total of $N$ agents (gray disks), each with radius $R$. The inset shows the initial configuration with the shortest agent-wall distance being $4R$. In (b), the pairwise velocity update from $(\vec{v}_i, \vec{v}_j)$ to $(\vec{v}_i', \vec{v}_j')$ incorporates elastic collisions, group cohesion, and social forces between agents $i$ and $j$. Additionally, agent-wall interactions are also taken into account. The update mechanisms indicated by the black starburst marker are listed in Table I.

Here, the reduced mass is

$$\mu_{ij} = \frac{m_i m_j}{m_i + m_j}, \tag{3}$$

the relative velocity is $\delta\vec{v}_{ij} \equiv \vec{v}_i - \vec{v}_j$, and the normal unit vector is

$$\hat{n}_{ij} = \frac{\delta\vec{r}_{ij}}{\delta r_{ij}} \tag{4}$$

with the relative position

$$\delta\vec{r}_{ij} = \vec{r}_i - \vec{r}_j, \tag{5}$$

where $\vec{r}_i$ and $\vec{r}_j$ are the position vectors at the moment of contact between agents $i$ and $j$. For agent-wall collisions, we consider hard boundaries where the velocity is updated according to

$$\vec{v}_{i,w}' = \vec{v}_i - (1+e)\,(\vec{v}_i \cdot \hat{n}_w)\,\hat{n}_w \tag{6}$$

with $\hat{n}_w$ being the unit vector normal to the corresponding wall. The boundary buffering mechanism utilizing $\vec{v}_{i,w}'$ is activated exclusively when an agent-agent collision occurs within a distance of $0.1L$ from a wall. Note that the inter-agent scattering, Eq. (2), conserves agent momentum, while the external forces due to the walls, Eq. (6), break this conservation. To examine effects governed solely by the proposed behavioral modulations, we set the coefficient of restitution to

$$e = 1, \tag{7}$$

ensuring elastic scattering where kinetic energy is conserved.

Finally, to account for the tendency of an agent to reorient their velocity toward nearby group members right after a collision, we model group cohesion as a net attractive direction acting on agent $i$ (belonging to group $g$) via the superposition of pairwise interactions:

$$\vec{v}'_{i,g} = k \sum_{j \neq i, \{i,j\} \in g} \frac{\vec{r}_{ij}}{|\vec{r}_{ij}|^{n+1}}. \tag{8}$$

Here, $n = 1$ is chosen so that nearby group members exert a stronger attractive pull than distant ones, while $k$ denotes a positive proportionality constant. Immediately following an agent-agent collision, agent $i$ undergoes a directional reorientation driven by the social-group cohesion. The collective influence of the three activated mechanisms—standard collision [Eq. (2)], group cohesion (8), and wall buffering (6)—determines the net post-collision velocity direction through a weighted sum, yielding the update

$$\vec{v}'_i = v'_{i,c} \hat{u}_i. \tag{9}$$

The unit vector

$$\hat{u}_i = \frac{\vec{w}_i}{|\vec{w}_i|} \tag{10}$$

is computed from the linear combination

$$\vec{w}_i = (1 - \gamma_w) \left[ (1 - \gamma_g) \hat{v}'_{i,c} + \gamma_g \hat{v}'_{i,g} \right] + \gamma_w \hat{v}'_{i,w}. \tag{11}$$

Here, $\hat{v}'_{i,c}$, $\hat{v}'_{i,g}$, and $\hat{v}'_{i,w}$ are unit vectors normalized from Eqs. (2), (8), and (6), respectively. Note that in Eq. (9), kinetic energy remains conserved because the updated velocity $\vec{v}'_i$ retains the post-collision speed magnitude $v'_{i,c}$ obtained from the standard elastic collision. In Eq. (11), $\gamma_g$ represents the strength of social-group cohesion, while $\gamma_w$ indicates that of wall buffering. The interaction weights relative to the standard elastic collision term are governed by these parameters, $\gamma_w, \gamma_g \in [0, 1]$, ensuring that the coefficients in $\vec{w}_i$ are nonnegative and sum to unity. For instance, in the limiting case where $\gamma_w = \gamma_g = 0$, the update simplifies to $\vec{v}'_i = \vec{v}'_{i,c}$, thereby recovering the baseline elastic collision model.

### B. Social force model (agent-agent and agent-wall forces)

To account for non-contact interactions, we adopt the SFM [6, 29]. Below, we address how this model is employed and interpret its components in the context of the present problem. In the SFM, the characteristic time scale is the reaction time $\delta t$, defined as the interval over which each agent updates its velocity according to

$$\vec{v}'_i = \vec{v}_i + \frac{\vec{F}_{tot}}{m_i} \delta t. \tag{12}$$

The total force

$$\vec{F}_{tot} = \vec{F}_i + \vec{F}_{i,\mathrm{wall}} \tag{13}$$

is assumed to act continuously in space and persistently in time. The agent-agent force $\vec{F}_i$ contribution is expressed as

$$\vec{F}_i = \vec{F}^d_i + \vec{F}^p_i + \vec{F}^s_i. \tag{14}$$

Here

$$\vec{F}^d_i = m_i \frac{\vec{v}^* - \vec{v}_i}{\delta t} \tag{15}$$

is the driving force, which relaxes agents toward the desired velocity $\vec{v}^*$. In the present setting, this desired motion may represent, for example, a tendency to move toward a focal location associated with the activity, such as a performance stage at a concert. We assume that this target lies at $x = L$, so that $\vec{v}^*$ points in the $+x$ direction. The driving force $\vec{F}^d_i$ is essential for sustaining active motion. In particular, if the desired velocity is set to zero, $\vec{v}^* = 0$, then the remaining two terms, namely the pushing force $\vec{F}^p_i$ and the sliding force $\vec{F}^s_i$, dissipate kinetic energy and suppress pairwise relative motion, as will be addressed below. This is the trivial case in which all agents eventually come to rest and no mechanically dangerous high-pressure state develops during the activity. Specifically, with agent radius $R$ and $\delta r_{ij} - 2R > 0$, the pushing contribution in the normal direction is

$$\begin{aligned} \vec{F}^p_i &= \sum_{j \neq i} \vec{F}^p_{ij} \\ &= \alpha m_i \sum_{j \neq i} \left( -\frac{\delta \vec{v}_{ij} \cdot \hat{n}_{ij}}{\delta t} \right) \exp \left( -\frac{\delta r_{ij} - 2R}{B} \right) \hat{n}_{ij}, \end{aligned} \tag{16}$$

while the sliding contribution in the tangential direction $\hat{t}_{ij} \equiv \hat{z} \times \hat{n}_{ij}$ is

$$\begin{aligned} \vec{F}^s_i &= \sum_{j \neq i} \vec{F}^s_{ij} \\ &= \beta m_i \sum_{j \neq i} \left( -\frac{\delta \vec{v}_{ij} \cdot \hat{t}_{ij}}{\delta t} \right) \exp \left( -\frac{\delta r_{ij} - 2R}{B} \right) \hat{t}_{ij}. \end{aligned} \tag{17}$$

We generalize these interactions to the non-contact regime by assuming that $\vec{F}^p_i$ and $\vec{F}^s_i$, with strengths $\alpha$ and $\beta$, decay exponentially with the agent-agent distance $\delta r_{ij}$ (with characteristic decay length $B$). From Eqs. (16) and (17), one has $\vec{F}^{p/s}_{ij} = -\vec{F}^{p/s}_{ji}$. Hence, the pairwise rate of kinetic-energy change (power),

$$\vec{v}_i \cdot \vec{F}^{p/s}_{ij} + \vec{v}_j \cdot \vec{F}^{p/s}_{ji} = \delta \vec{v}_{ij} \cdot \vec{F}^{p/s}_{ij}, \tag{18}$$

is non-positive. In particular, for the pushing term,

$$\delta\vec{v}_{ij} \cdot \vec{F}_{ij}^{p} \propto -\left(\delta\vec{v}_{ij} \cdot \hat{n}_{ij}\right)^2 \le 0, \tag{19}$$

whereas for the sliding term,

$$\delta\vec{v}_{ij} \cdot \vec{F}_{ij}^{s} \propto -\left(\delta\vec{v}_{ij} \cdot \hat{t}_{ij}\right)^2 \le 0. \tag{20}$$

Therefore, both $\vec{F}^p$ and $\vec{F}^s$ are friction-like, energy-dissipating terms. Specifically, $\vec{F}^p$ reduces the approach speed between nearby agents, while $\vec{F}^s$ damps relative tangential motion to suppress lateral slip between neighboring agents.

The walls are treated as obstacles by the agents, and we introduce analogous pushing and sliding as

$$\vec{F}_{i,w}^{p} = \alpha_w m_i \left(-\frac{\vec{v}_i \cdot \hat{n}_w}{\delta t}\right) \exp\left(-\frac{\delta r_{i,w} - R}{B}\right) \hat{n}_w \tag{21}$$

and

$$\vec{F}_{i,w}^{s} = \beta_w m_i \left(-\frac{\vec{v}_i \cdot \hat{t}_w}{\delta t}\right) \exp\left(-\frac{\delta r_{i,w} - R}{B}\right) \hat{t}_w, \tag{22}$$

with $\hat{t}_w \equiv \hat{z} \times \hat{n}_w$. Here, $\alpha_w$ and $\beta_w$ denote the strengths of the agent-wall pushing and sliding terms, $\vec{F}_{i,w}^{p}$ and $\vec{F}_{i,w}^{s}$, respectively. The quantity $\delta r_{i,w}$ represents the normal distance from agent $i$ to the corresponding wall. By summing over all walls $w = l, r, b, t$, we compute the agent-wall force contribution as

$$\vec{F}_{i,\text{wall}} = \sum_{w} \left(\vec{F}_{i,w}^{p} + \vec{F}_{i,w}^{s}\right). \tag{23}$$

## III. RESULTS AND DISCUSSION

In this section, we present and analyze the results of our numerical simulations. Unless otherwise specified, the following default parameters are adopted. The system comprises $N = 400$ agents, each modeled as a rigid disk of mass $m = 60$ kg and radius $R = 0.25$ m. The agents are confined within a square domain of side length $L = 31.25$ m. For the SFM dynamics, the reaction time is set to $\delta t = 0.45$ s, and the decay length of the forces is $B = 1.5R$. The macroscopic time step $\Delta t = 6$ s in Eq. (1) is selected. Default MKS units are used in our figures unless specified otherwise. Before proceeding to the main findings, we first demonstrate the robustness of our results against the choice of the ICs. Although a uniform initial distribution is strictly employed in the subsequent analyses (Secs. III A, III B, and III C), we establish here that the steady-state macroscopic metric $P_{\text{wall}}$ is independent of this specific choice.

We consider three distinct distribution $f(v)$ types: the uniform, the Maxwell-Boltzmann, and a truncated power-law distribution (proportional to $v^{-h}$). The uniform distribution is given by

$$f_u(v) = \frac{\theta(v_{\max} - v)}{v_{\max}}, \tag{24}$$

which yields the root-mean-square speed,

$$v_{\max} = \sqrt{3} v_{rms} \tag{25}$$

and $\theta(\cdot)$ denotes the Heaviside unit step function. The Maxwell-Boltzmann distribution takes the form

$$f_{MB}(v) = \frac{v}{\sigma^2} \exp\left(-\frac{v^2}{2\sigma^2}\right), \tag{26}$$

which is parameterized by $v_{rms}$ via the relation

$$v_{rms} = \sqrt{2}\sigma. \tag{27}$$

For comparison, we introduce a hypothetical bounded distribution obeying the power-law (with $1 < h < 3$) as

$$f_h(v) = \frac{h-1}{v_{\min}^{1-h} - v_{\max}^{1-h}} v^{-h} \times \theta(v - v_{\min}) \theta(v_{\max} - v), \tag{28}$$

yielding the root-mean-square speed

$$v_{rms} = \sqrt{\frac{h-1}{3-h} \times \frac{v_{\max}^{3-h} - v_{\min}^{3-h}}{v_{\min}^{1-h} - v_{\max}^{1-h}}}. \tag{29}$$

All three distributions satisfy the normalization condition

$$\int_0^{\infty} f(v) dv = 1, \tag{30}$$

and are parameterized to share the same root-mean-square speed of

$$v_{rms} = 1.2 \text{ m/s}, \tag{31}$$

governed by Eqs. (25), (27), and (29), for $f_u(v)$, $f_{MB}(v)$, and $f_h(v)$, respectively. For the numerical implementation, we sample the speed domain using 25 discrete grid points. The number of agents $N_I$ assigned to the $I$-th speed bin is determined by

$$N_I = \mathcal{R}\left(N \times \frac{f(v_I)}{Z}\right), \tag{32}$$

where $\mathcal{R}(\cdot)$ denotes the rounding function, and

$$Z = \sum_{I=1}^{25} f(v_I) \tag{33}$$

is the normalization constant. As an example, for the power-law distribution $f_h(v)$, we adopt an exponent $h = 2.5$ and velocity bounds $(v_{\min}, v_{\max}) = (0.48, 6.13)$ m/s. Under the present discretization, this parameterization yields the designed root-mean-square speed, Eq. (31), whereas the theoretical continuous limit evaluated via Eq. (29) yields $v_{rms} \approx 1.36$ m/s.

For a fixed root-mean-square speed $v_{\text{rms}}$ Eq. (31), Fig. 2 illustrates the dynamics of the baseline elastic collision model evolving from three distinct initial speed distributions: $f_u(v)$ [Figs. 2(a)–2(c)], $f_{\text{MB}}(v)$ [Figs. 2(d)–2(f)], and $f_h(v)$ [Figs. 2(g)–2(i)]. Although the initial distributions in Figs. 2(a), 2(d), and 2(g) are highly distinct, their final distributions, at

$$t_f \approx 11.16 \text{ minutes} \tag{34}$$

well after $P_{\text{wall}}$ becomes steady, in Figs. 2(b), 2(e), and 2(h) share two key features: $(i)$ a long-tailed Maxwell-Boltzmann-like profile and $(ii)$ the same most-probable speed. While instantaneous temporal fluctuations arise due to the finite system size ($N = 400$), these overall statistical features remain unchanged. More importantly, all distributions converge to the same steady macroscopic wall line pressure $P_{\text{wall}}$ [Eq. (1)]. This is evidenced by the identical macroscopic slopes of the accumulated momentum $p_w(t) = \sum \delta p_w$ in the insets of Figs. 2(c), 2(f), and 2(i), where we also verify total kinetic energy conservation. Although the microscopic momentum transfer $\delta p_w$ undergoes frequent spikes owing to shot noise from discrete agent-wall collision events, the system reaches a steady $P_{\text{wall}}$ in a short time ($t \approx 16$ s). By this time, numerous collisions (on the order of $10^3$) have occurred due to the dense crowding of the agents.

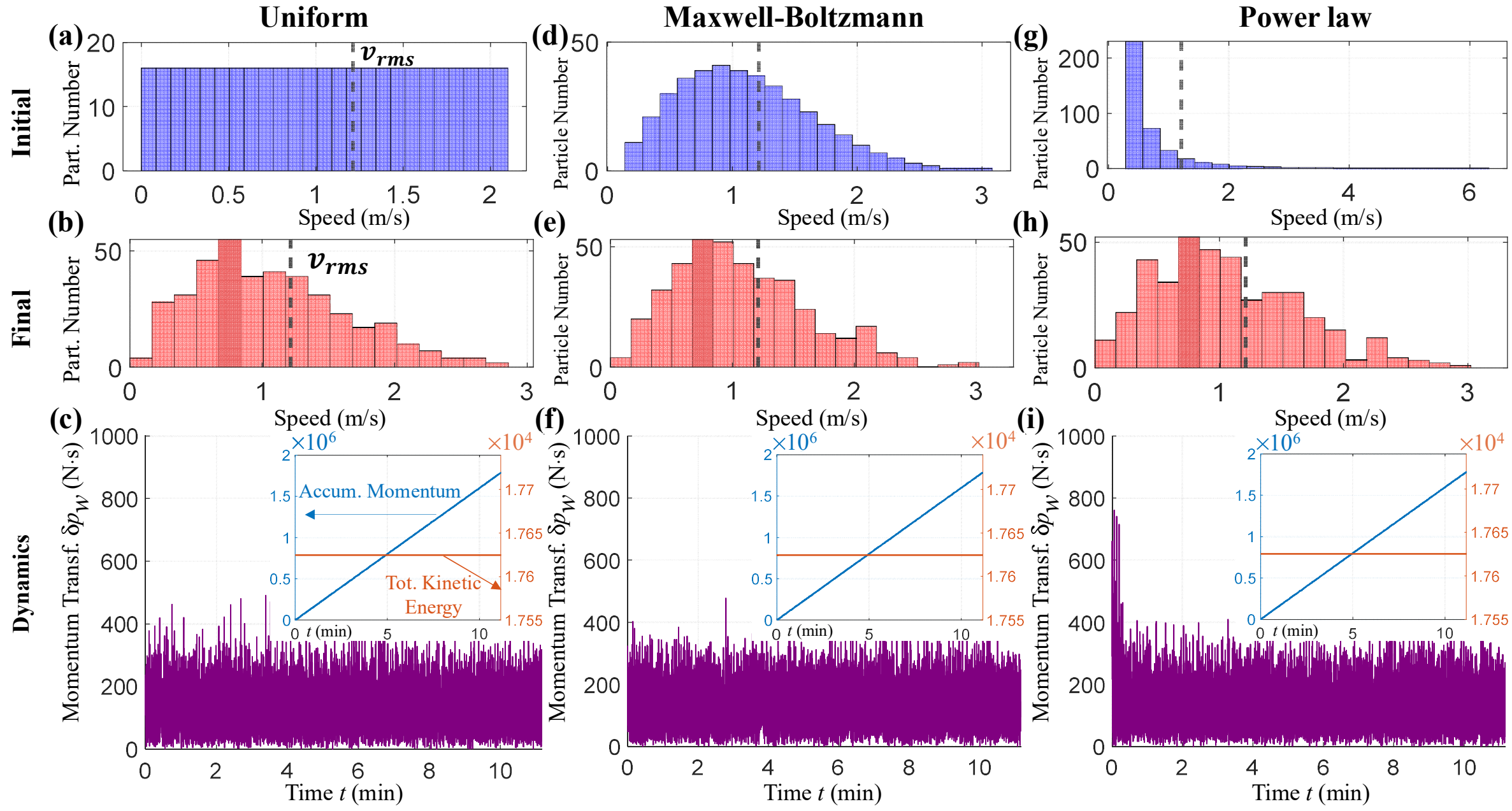

FIG. 2. Dynamics of the baseline elastic collision model for uniform [(a)–(c)], Maxwell-Boltzmann [(d)–(f)], and power-law [(g)–(i)] initial distributions. The first row [(a), (d), and (g)] shows the initial binned distributions generated according to Eq. (32), which evolve into the final distributions shown in the second row [(b), (e), and (h)] at $t_f = 11.6$ minutes, well after $P_{\text{wall}}$ becomes steady. The vertical dashed lines indicate the root-mean-square speed $v_{rms}$, while the dark red (emphasized) bars indicate the most-probable speed. The third row exhibits the time-resolved impulse $\delta p_w$ due to agent-wall collisions. The insets show the accumulated momentum (left axis), where the coarse-grained slopes, Eq. (1), yield the macroscopic metric $P_{\text{wall}}$; additionally, total kinetic energy conservation is verified (right axis).

For the system with all interaction layers in Table I enabled, with $v_{\text{rms}}$ fixed according to Eq. (31) and $v^* = v_{\text{rms}}$, Fig. 3 exhibits a Maxwell-Boltzmann-like final profile similar to that in Fig. 2. However, unlike the baseline elastic collision model, total kinetic energy is no longer conserved due to the active social forces. Nonetheless, the steady wall line pressure $P_{\text{wall}}$ at $t_f$ [Eq. (34)] is identical and independent of the initial distribution. This independence underscores the generality of the $P_{\text{wall}}$ features analyzed below. Here, $t_f$ in practice corresponds to the time at which evacuation commences, i.e., the exit remains unavailable until $t_f$. Conversely, the maximum per-agent contact load $\delta p_{\text{max}}$ recorded before $t_f$ depends on the initial distribution. As an extreme example, for a given $v_{\text{rms}}$, if only one agent possesses a nonzero initial speed while all others are at rest, then $\delta p_{\text{max}}$ is already maximized at $t = 0$. In fact, our numerical results suggest that the observed $\delta p_{\text{max}}$ approaches this theoretical limit as $t_f$ increases. In light of this, below we employ a uniform distribution to provide a mild initial $\delta p$ and focus on the developed $\delta p_{\text{max}}$ prior to $t_f$. Note that because $\delta p_{\text{max}}$ reflects both agent-agent and agent-wall collisions, a larger $P_{\text{wall}}$ does not necessarily imply a larger $\delta p_{\text{max}}$.

### A. Kinetic-energy-conserving reorientation

For a given total number of agents $N$, we examine the role of $\gamma_g$ in determining $P_{\text{wall}}$ and $\delta p_{\text{max}}$ within the elastic reorientation model (ERM), where both the social-group cohesion layer $\gamma_g$ and the wall buffering $\gamma_w$ mechanism in Table I are activated. The set $\mathbf{N} = \{N_0; N_{g=1}, N_{g=2}, \cdots, N_{g=G}\}$ denotes the group configuration, where $N_0$ is the number of independent agents (those not belonging to any group), $N_g$ is the size of the $g$-th group, and $G$ is the total number of groups. Figure 4 shows $P_{\rm wall}$ in panels (a1)–(f1) and $\delta p_{\max}$ in panels (a2)–(f2) as a function of $\gamma_w$ and $\gamma_g$. In Fig. 4, across the top row, the number of independent agents is reduced via $N_0 = 400$, 200, and 0 for columns (a), (b), and (c), respectively; conversely, the bottom row explores the fully grouped limit $N_0 = 0$, where the total number of groups decreases as $G = 200$, 20, and 1 for columns (d), (e), and (f), respectively.

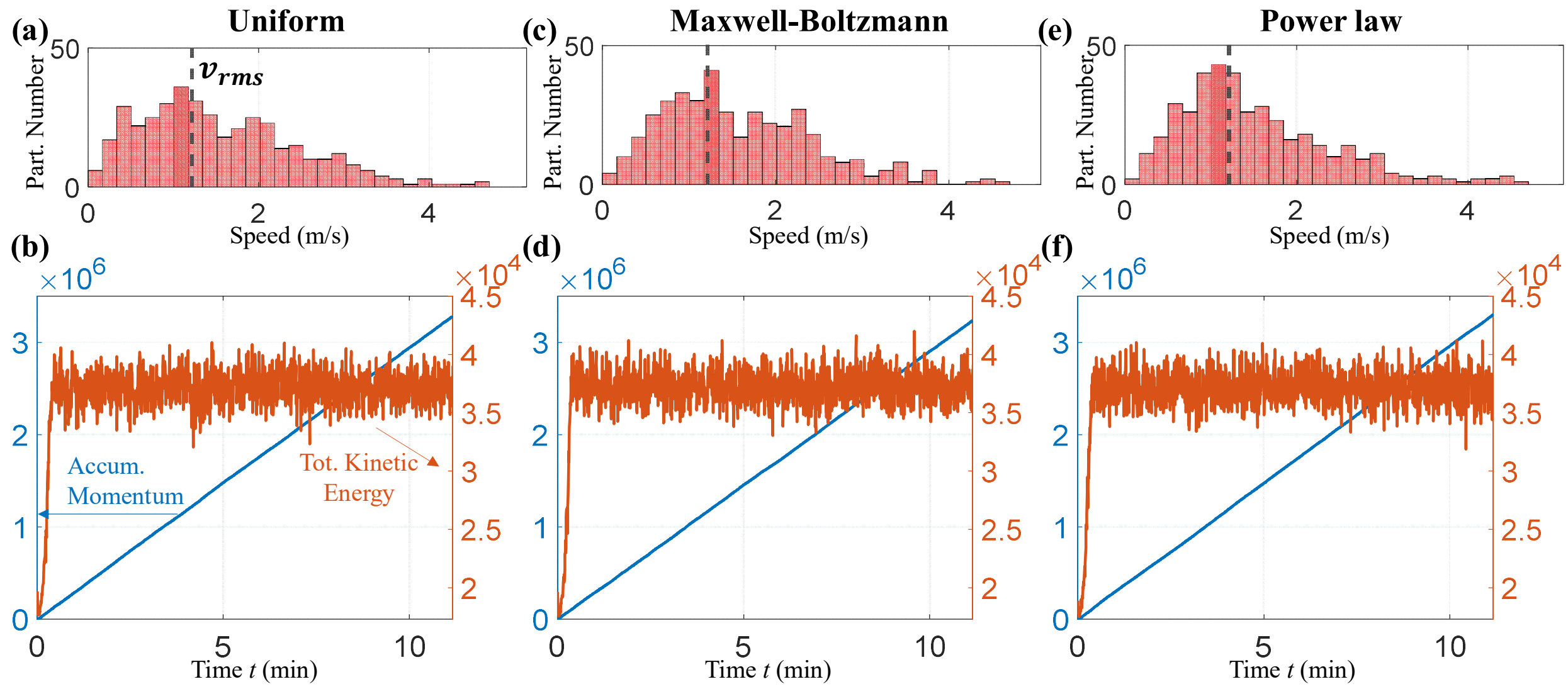

FIG. 3. Dynamics with all interaction layers enabled, starting from the same initial conditions as in Fig. 2, for uniform [(a) and (b)], Maxwell-Boltzmann [(c) and (d)], and power-law [(e) and (f)] initial distributions. The first row shows the final speed distributions, with dark red bars indicating the most-probable speed. The second row shows the time-resolved accumulated wall momentum $p_w(t)$ and the total kinetic energy. Interaction strengths $\gamma_w = \gamma_g = 0.5$ are used, with the system divided into two groups of 200 agents each.

In the non-grouped/all-independent limit $N_0 = N$, shown in Figs. 4(a1) and 4(a2), $\gamma_g$ has no effect on the dynamics because none of the agents are grouped. In this regime, $P_{\rm wall}$ is suppressed solely by the wall-buffering mechanism $\gamma_w$. As agents are gradually assigned to groups, $\gamma_g$ begins to play an important role in mitigating wall pressure [compare Figs. 4(a1), (b1), and (c1)]. In particular, high values of $P_{\rm wall}$ are identified in the lower-left region of small $\gamma_w$ and $\gamma_g$ satisfying

$$\begin{aligned} \Gamma(\gamma_w, \gamma_g) &\equiv (1-\gamma_w)(1-\gamma_g), \\ \Gamma(\gamma_w, \gamma_g) &\geq 0.5, \end{aligned} \tag{35}$$

with the maximum of $P_{\rm wall}$ occurring at the baseline elastic limit $\gamma_w = \gamma_g = 0$. Furthermore, the reduction of $P_{\rm wall}$ via $\gamma_w$ corresponds microscopically to a smaller evolved population of near-wall agents (i.e., most agents remain within the bulk). This characteristic is clearly visible in the spatial distributions Fig. 5, where the dashed (solid) markers denote lower (higher) values of $\gamma_w$ selected from Fig. 4. In the limit of the largest group size ($G = 1$ or $N_1 = N$), the optimal mitigation of $P_{\rm wall}$ occurs around $\gamma_g \lesssim 0.5$. According to Eq. (11), this characteristic value represents a near-balanced weighting of standard elastic scattering and group-cohesive reorientation, with a slight bias toward the standard mechanism. Conversely, as a general strategy for mitigating $P_{\rm wall}$, effective, though not optimal, reduction is achieved in any configuration with a sizable group (large $N_g$) when agents favor group cohesion ($\gamma_g > 0.5$). This $\gamma_g$-driven pressure mitigation stems directly from the formation of group clustering within the bulk, as depicted by the solid circle and solid triangle markers in Figs. 5(b) and 5(c), which lead to suppressed agent-wall collisions.

We now discuss the microscopic $\delta p_{\max}$. The high $P_{\rm wall}$ observed in the small-$\gamma$ regime [Eq. (35)] is not reflected in $\delta p_{\max}$. In particular, $\delta p_{\max}$ in Fig. 4 does not show a strong dependence on $\gamma_w$, implying that it is determined mainly by bulk group-clustering dynamics. Specifically, we find that $\delta p_{\max}$ is governed by clustered intra-group agent-agent collisions. Intuitively, a larger $\gamma_g$ drives a stronger tendency to form clusters, which would be expected to result in a larger $\delta p_{\max}$. However, we observe that a high-$\delta p_{\max}$ hazard window emerges at intermediate $\gamma_g \approx 0.6 \pm 0.1$ as the group size $N_g$ increases [see the progression from Figs. 4(a2) to 4(c2) and from 4(d2) to 4(f2)]. In the limit of the largest group size Fig. 4(f2), this hazard window is most prominent. In contrast, the extreme case of $\gamma_g = 1$ [where group cohesion completely eliminates the standard elastic collision component in Eq. (11)] produces a relatively low $\delta p_{\max}$. To understand this counterintuitive suppression of $\delta p_{\max}$ by a very large $\gamma_g \approx 1$, we provide an analysis below based on the cluster size and cluster center-of-mass (CCM) kinetic energy.

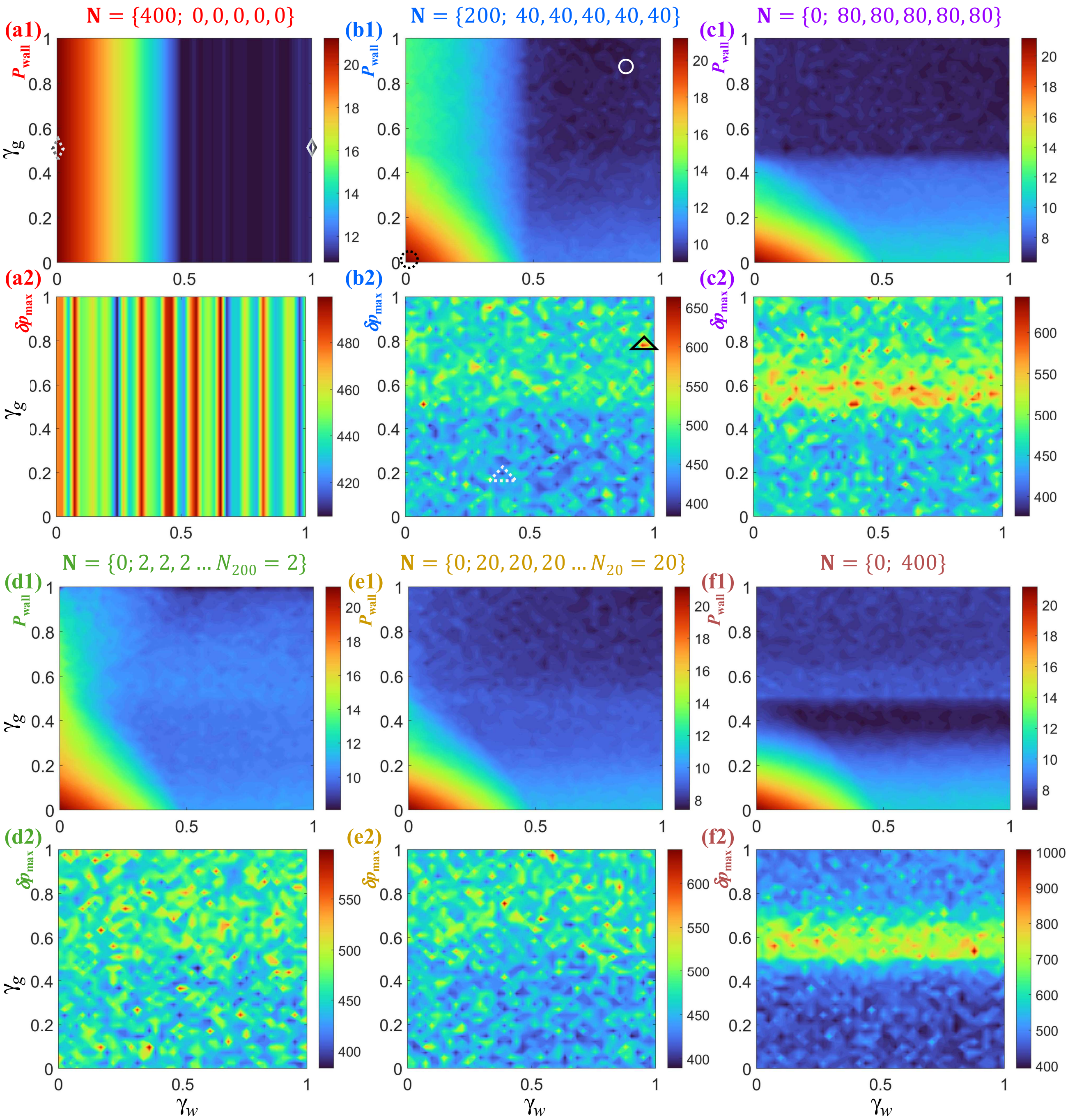

FIG. 4. (a1)–(f1) $P_{\mathrm{wall}}$ and (a2)–(f2) $\delta p_{\max}$ as functions of $\gamma_w$ and $\gamma_g$ in the elastic reorientation model (ERM) under different group configurations $\mathbf{N} = \{N_0; N_{g=1}, N_{g=2}, \cdots, N_{g=G}\}$. Across the top row, the number of independent agents is reduced via $N_0 = 400$, 200, and 0 for columns (a), (b), and (c), respectively. The bottom row adopts the fully grouped limit $N_0 = 0$, with decreasing $G = 200$, 20, and 1 for columns (d), (e), and (f), respectively. The dashed and solid markers in (a1), (b1), and (b2) mark the representative cases selected for Fig. 5, with the corresponding $\gamma$ values listed in its bottom-left panel.

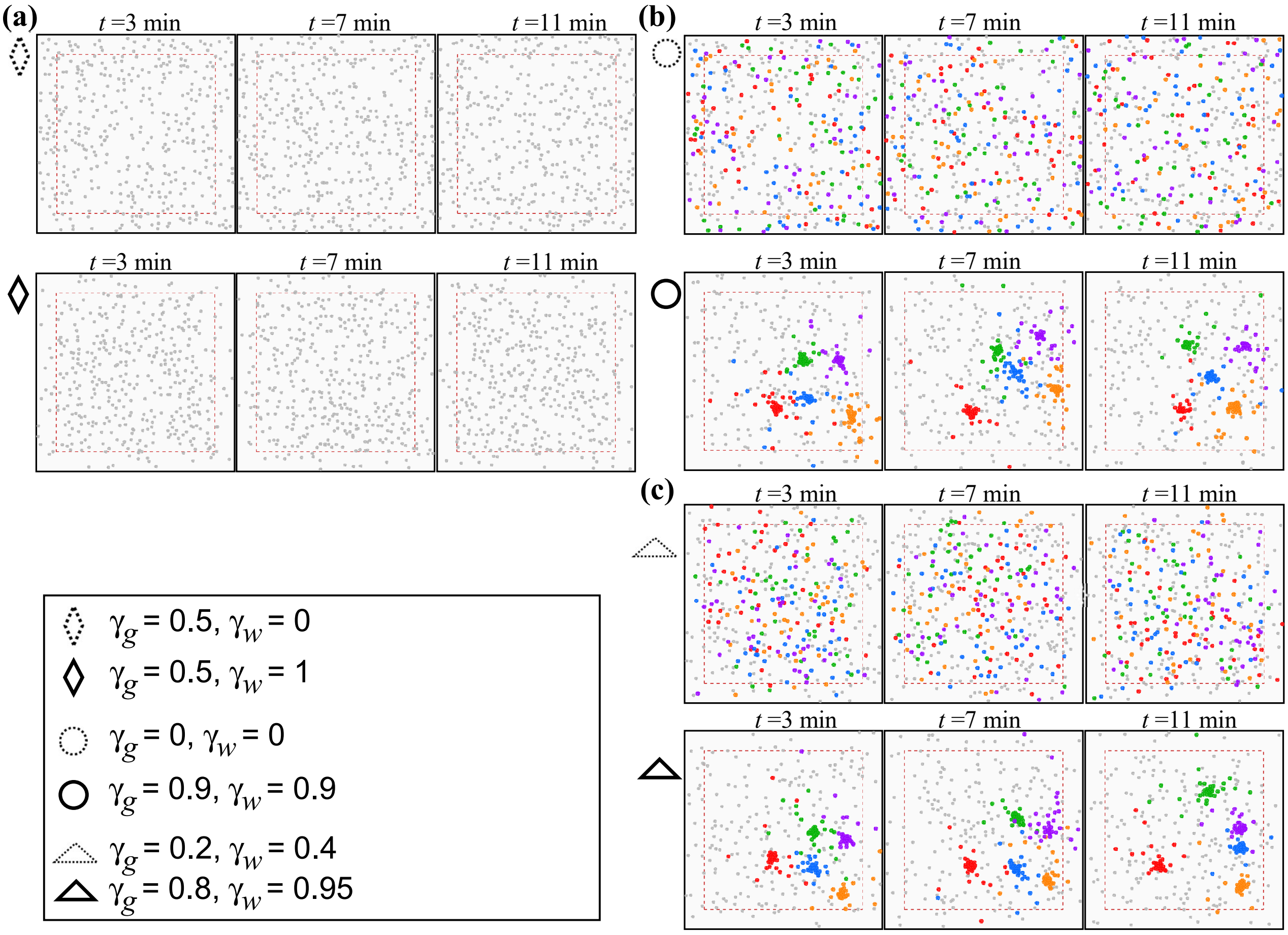

FIG. 5. Representative snapshots of agent spatial distributions at $t = 3$, 7, and 11 min, with the dashed and solid markers selected from Fig. 4; the corresponding $\gamma$ values are listed in the bottom-left panel. In the non-grouped/all-independent case (a) and the grouped case (b), dashed (solid) markers correspond to the high (low) $P_{\rm wall}$. In (c), dashed (solid) markers correspond to the low (high) $\delta p_{\max}$. Agents in the same group are shown by dots of the same color, whereas gray dots denote independent agents (those not belonging to any group). The dashed square marks the region within which wall-buffering $\gamma_w$ is effective.

Consider the regime in which group cohesion dominates standard elastic collisions, especially in the limit $\gamma_g \to 1$. Since the cohesion strength decays as $1/r_{ij}$ [Eq. (8) with $n = 1$], if two agents $A$ and $B$ belonging to the same group $g$ are far from the other members of their group, after a collision they act as if bound together, with the $A$-$B$ center-of-mass moving only locally. We refer to this behavior of $A$ and $B$ as the local pairing effect, which introduces distinct features. First, if both agents are far from the main $g$-cluster, this local pairing restricts their ability to explore the bulk space and join the main $g$-cluster. Second, if agent $A$ is already clustered on the fluctuating, sparse surface of the $g$-cluster, an intra-group collision causes $A$ to decouple from the cluster's collective motion. These features become increasingly pronounced at large group sizes $N_g$ [e.g., Fig. 4(f2)], so that the cluster size at $\gamma_g \approx 1$ is smaller than that observed in the mild cohesion regime ($\gamma_g \approx 0.6$). Furthermore, the total kinetic energy of a cluster, $K^c$, can be decomposed as

$$K^c = K^c_{CM} + K^c_{rel}, \tag{36}$$

where $K^c_{CM}$ arises from the CCM motion and $K^c_{rel}$ stems from the relative motion of the constituent agents. Importantly, $K^c_{rel}$ drives the clustered intra-group agent-agent collisions that dictate $\delta p_{\max}$, whereas $K^c_{CM}$ does not contribute to contact loads. The total cluster kinetic energy $K^c$ remains approximately constant due to elastic intra-group collisions. We find that $K^c_{CM}$ is substantially greater at $\gamma_g \approx 1$ than at $\gamma_g \approx 0.6$; for example, in Fig. 4(f2), $K^c_{CM}$ near $\gamma_g = 1$ exceeds its value at $\gamma_g = 0.6$ by a factor of approximately 1.7. In other words, when approaching the extreme $\gamma_g = 1$ at a large $N_g$, not only do fewer agents remain clustered, but a smaller fraction of kinetic energy is allocated to $K^c_{rel}$, eventually resulting in a smaller $\delta p_{\max}$.

In practice, small $N_g$ helps keep $\delta p_{\max}$ low; when $N_g$ is large, avoiding the hazard window of $\delta p_{\max}$ requires agents to avoid a balanced choice between group cohesion and standard collision. Since $\delta p_{\max}$ captures the maximum impulse from both agent-agent and agent-wall collisions, our ERM results in Fig. 4 suggest the following overall strategy: agents first prioritize mitigating $\delta p_{\max}$ through a standard-collision-biased update with $\gamma_g < 0.5$; then, if a low $P_{\text{wall}}$ is also desired, agents adopt a strong wall-buffering-based update with $\gamma_w > 0.5$.

### B. Social force model interactions

Taking into account realistic inelastic scattering within the SFM framework, we analyze how social forces alone affect $P_{\text{wall}}$ and $\delta p_{\max}$. The results are shown in Fig. 6 and are independent of how the agents are configured owing to the absence of the $\gamma$ layers. The desired velocity $\vec{v}^* = v_{rms}\hat{x}$ in the active driving force $\vec{F}^d_i$ [Eq. (15)] results in agent clogging near the $x = L$ wall, where most of the kinetic energy accumulates. As a consequence, an approximate $\sqrt{N}$ enhancement of $P_{\text{wall}}$ emerges, corresponding to the bulk-to-surface ratio $N/\sqrt{N} = 20$, compared with the case without $\vec{F}^d_i$ [e.g., compare Fig. 6(a1) with Fig. 4(a1)]. Comparing agent-agent pushing $\alpha$ with sliding $\beta$ and agent-wall pushing $\alpha_w$ with wall sliding $\beta_w$, a reduction in $P_{\text{wall}}$ is primarily driven by sliding dominating over pushing ($\beta$ over $\alpha$) and wall pushing dominating over wall sliding ($\alpha_w$ over $\beta_w$). However, these parameters only slightly modulate $P_{\text{wall}}$, as seen in Figs. 6(a1)–(c1). Conversely, as shown in Figs. 6(a2)–(c2), $\delta p_{\max}$ is significantly increased by both $\alpha$ and $\beta$. In addition to $\alpha$ and $\beta$, $\delta p_{\max}$ is *further magnified* by the coexistence of $\alpha_w$ and $\beta_w$. These results suggest that pushing and sliding should be avoided in order to prevent high $\delta p_{\max}$, thereby improving crowd safety by maintaining a low per-agent contact load.

### C. Coupled reorientation and social force model interactions

When all interaction layers are enabled in the combined ERM and SFM framework, it becomes difficult to identify a $(\gamma_w, \gamma_g)$ that minimizes both $P_{\text{wall}}$ and $\delta p_{\max}$ simultaneously. As a feature arising from the integration of the $\gamma$ layers with the SFM interactions, a low (high) $P_{\text{wall}}$ is often associated with a high (low) $\delta p_{\max}$. We refer to this behavior as the wall-pressure/contact-load trade-off ($P$-$p$ trade-off). This trade-off is shown in Fig. 7, whereas the ERM-only (or social-force-free) results in Fig. 4 do not generally exhibit the same inverse correspondence between $P_{\text{wall}}$ and $\delta p_{\max}$.

To understand the physical mechanism underlying the $P$-$p$ trade-off, we examine the distinct roles of the ERM and SFM parameters. In the ERM, $\gamma_w$ and $\gamma_g$ tend to retain agents within the bulk region; i.e., they concentrate kinetic energy in the bulk, where the maximum per-agent load $\delta p_{\max}$ is recorded. Conversely, the agent-agent social pushing $\alpha$ and sliding $\beta$ in the SFM tend to drive agents away from the bulk, concentrating kinetic energy near the walls, which accounts for the steady macroscopic wall pressure $P_{\text{wall}}$. Consequently, the coexistence of the ERM and SFM layers provides competing mechanisms for redistributing kinetic energy between the bulk and boundary regions, thereby giving rise to the trade-off between $P_{\text{wall}}$ and $\delta p_{\max}$. Note also that the agent-wall social forces $\alpha_w$ and $\beta_w$ collectively counteract agent accumulation at the boundaries. Thus, their isolated effects within the SFM framework act analogously to the bulk-retaining mechanisms of $\gamma_w$ and $\gamma_g$, leaving the basic structure of the $P$-$p$ trade-off unaltered in the combined ERM+SFM regime.

In Fig. 7, similar to the SFM-only case, $P_{\text{wall}}$ is enhanced by an approximate factor of $\sqrt{N}$ due to the active driving force $F^d$. Within this combined ERM+SFM regime, we also observe that for fully grouped agents $N_0 = 0$, the peak value of $\delta p_{\max}$ across the entire parameter space $(\gamma_w, \gamma_g) \in [0,1]^2$ increases noticeably with the group size $N_g$. For example, as $N_g$ scales from $N_g = 2$ to $N_g = 400$ [corresponding respectively to Figs. 7(d2), 7(e2), 7(c2), and 7(f2)], the maximum encountered $\delta p_{\max}$ escalates from approximately 650 N·s to 880 N·s. While the SFM assists the ERM in establishing the $P$-$p$ tradeoff, the $P_{\text{wall}}$ profiles due to the reorientation by $\gamma_w$ and $\gamma_g$ are mostly preserved. In Figs. 7(a1) and 7(a2), the independent agents $N_0$ are governed exclusively by $\gamma_w$ rather than $\gamma_g$, exhibiting distinct vertical stripes. These vertical stripes are absent in Figs. 7(c1)–7(f1) due to the complete absence of independent agents $N_0 = 0$. Notably, for $\gamma_w > 0.5$, a low $P_{\text{wall}}$ [similar to Figs. 4(a1) and 4(b1)] emerges but now becomes nearly constant, $P_{\text{wall}} \approx$ const. When groups are present ($G \geq 1$), the lower-left region of small $\gamma_w$ and $\gamma_g$ satisfying Eq. (35) again exhibits an elevated $P_{\text{wall}}$. We refer to the equality condition $\Gamma(\gamma_w, \gamma_g) = 0.5$ as the equality locus (EL). Motivated by the EL and the threshold $\gamma_w = 0.5$, we analyze the phase diagram below.

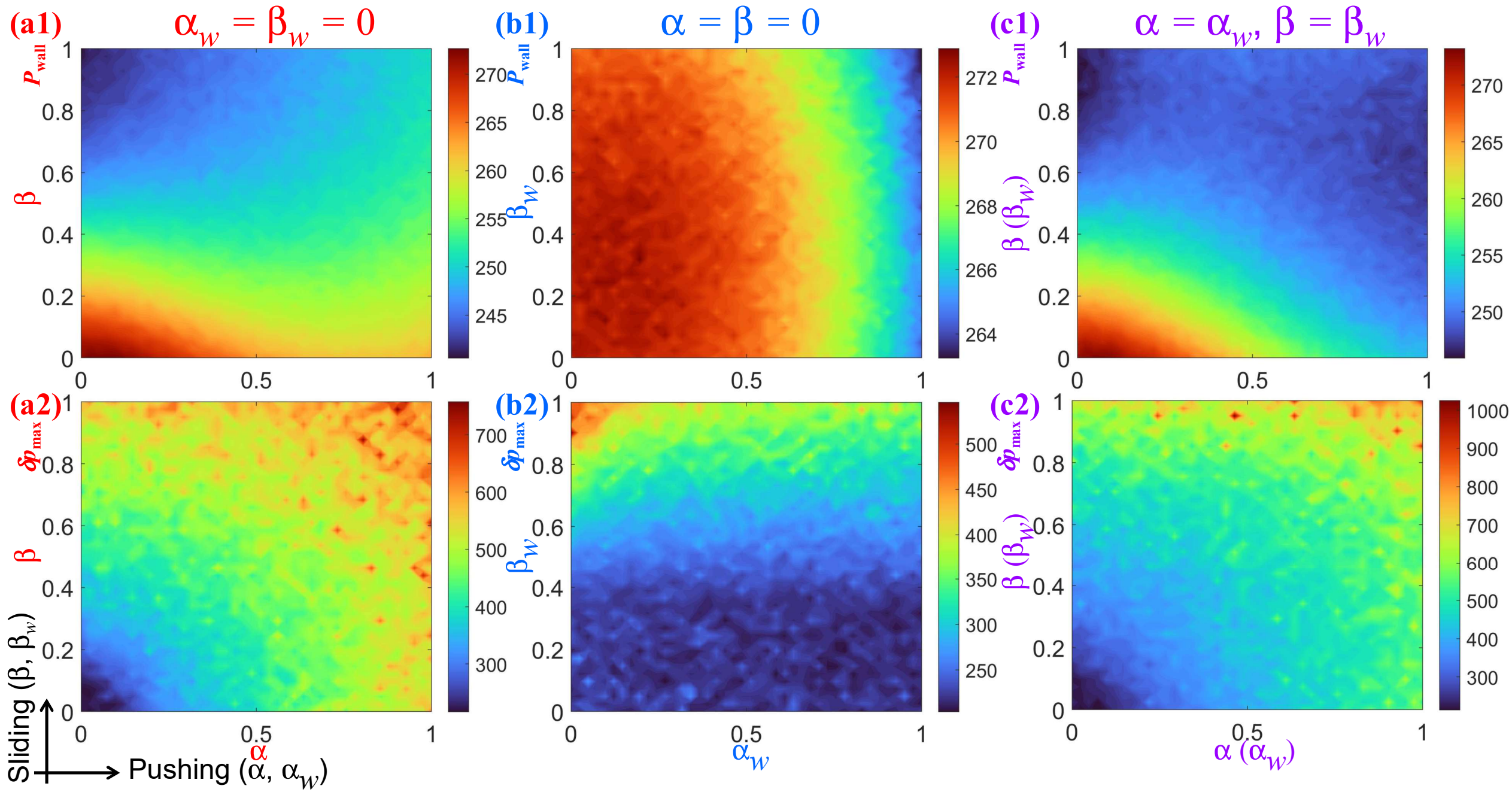

FIG. 6. (a1)–(c1) $P_{\text{wall}}$ and (a2)–(c2) $\delta p_{\max}$ as functions of the SFM pushing and sliding interactions in the presence of the driving force (15) with $\vec{v}^* = v_{rms}\hat{x}$. Column (a) isolates the agent-agent interactions with $\alpha_w = \beta_w = 0$, column (b) isolates the agent-wall interactions with $\alpha = \beta = 0$, and column (c) shows the coexistence case with $\alpha = \alpha_w$ and $\beta = \beta_w$.

To formally establish the macroscopic $P_{\text{wall}}$ phase regimes and determine whether they constitute genuine thermodynamic phase transitions, we perform a finite-size scaling analysis by increasing the total number of agents $N$ while keeping the areal density $N/L^2$ fixed. Here, the limit $N \to \infty$ corresponds to the thermodynamic limit. The susceptibility is defined as

$$\vec{\chi} = (\chi_w, \chi_g) \equiv (\partial P_{\text{wall}}/\partial \gamma_w, \partial P_{\text{wall}}/\partial \gamma_g). \tag{37}$$

The upper subpanels of Figs. 8(a)–(c) show the susceptibility curves obtained by scanning $\gamma_w$ [as indicated by the horizontal dashed lines with heart and club markers in Figs. 7(a1) and 7(b1), respectively] and $\gamma_g$ [as indicated by the vertical dashed line with the spade marker in Fig. 7(b1)]. The associated lower subpanels display the finite-size dependence of the susceptibility dip magnitudes, $|\chi_w^*|$ or $|\chi_g^*|$, extracted as the absolute value of the most negative susceptibility observed in the respective upper subpanels, along with the power-law fits $|\chi_w^*| \propto N^f$ and $|\chi_g^*| \propto N^f$.

In the $\gamma_w$ scans at $\gamma_g = 0.1$ in Figs. 8(a)–8(b), $\chi_w$ displays a susceptibility discontinuity $\Delta\chi_w$ at $\gamma_w = 0.5$, governed by the independent-agent fraction $N_0/N$. When $N_0 > 0$, the slope of this step-like jump diverges, $\partial\chi_w/\partial\gamma_w \to \infty$, in the thermodynamic limit. This susceptibility discontinuity diminishes as $N_0/N$ decreases and vanishes at $N_0 = 0$, where the susceptibility profile softens and $\gamma_w = 0.5$ ceases to define a phase boundary. Beyond this threshold, $\gamma_w > 0.5$, $P_{\text{wall}}$ plateaus ($P_{\text{wall}} \approx$ const.), demonstrating that strong individual wall awareness effectively decouples the wall pressure from internal crowd dynamics. The phase boundary at $\gamma_w = 0.5$ is depicted by the blue dashed line in Fig. 8(d).

We now turn to the behavior near the EL. For $\gamma_w = 0.1$, the EL intersects the $\gamma_g$ scan in Fig. 8(c) at $\gamma_g^* \approx 0.444$.

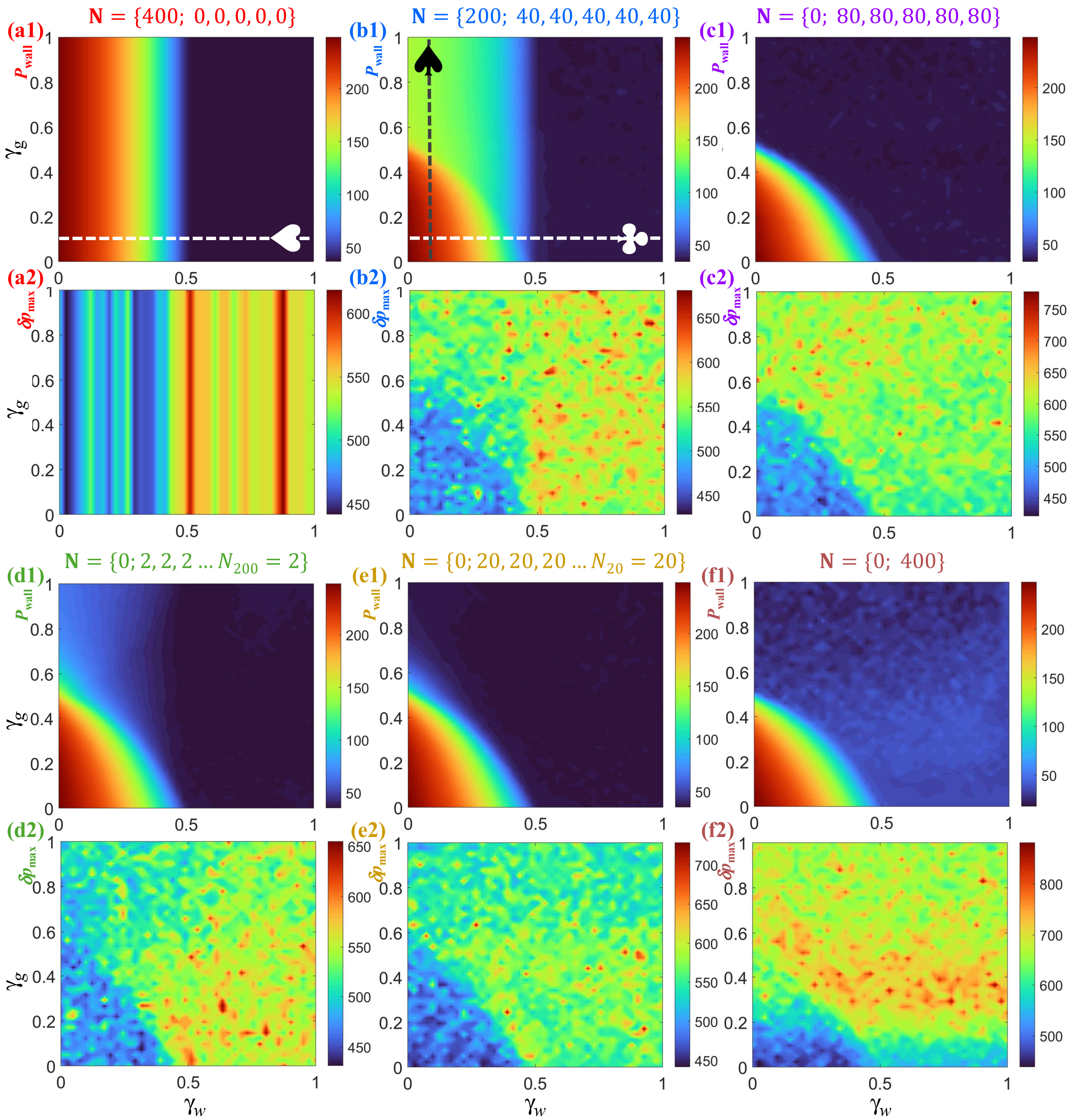

FIG. 7. $P_{\text{wall}}$ and $\delta p_{\max}$ as functions of $\gamma_w$ and $\gamma_g$ in the ERM+SFM framework with $\alpha = \beta = \alpha_w = \beta_w = 0.5$, under the same group configurations as in Fig. 4. The dashed lines with markers indicate the scans of $P_{\text{wall}}$ analyzed in Fig. 8.

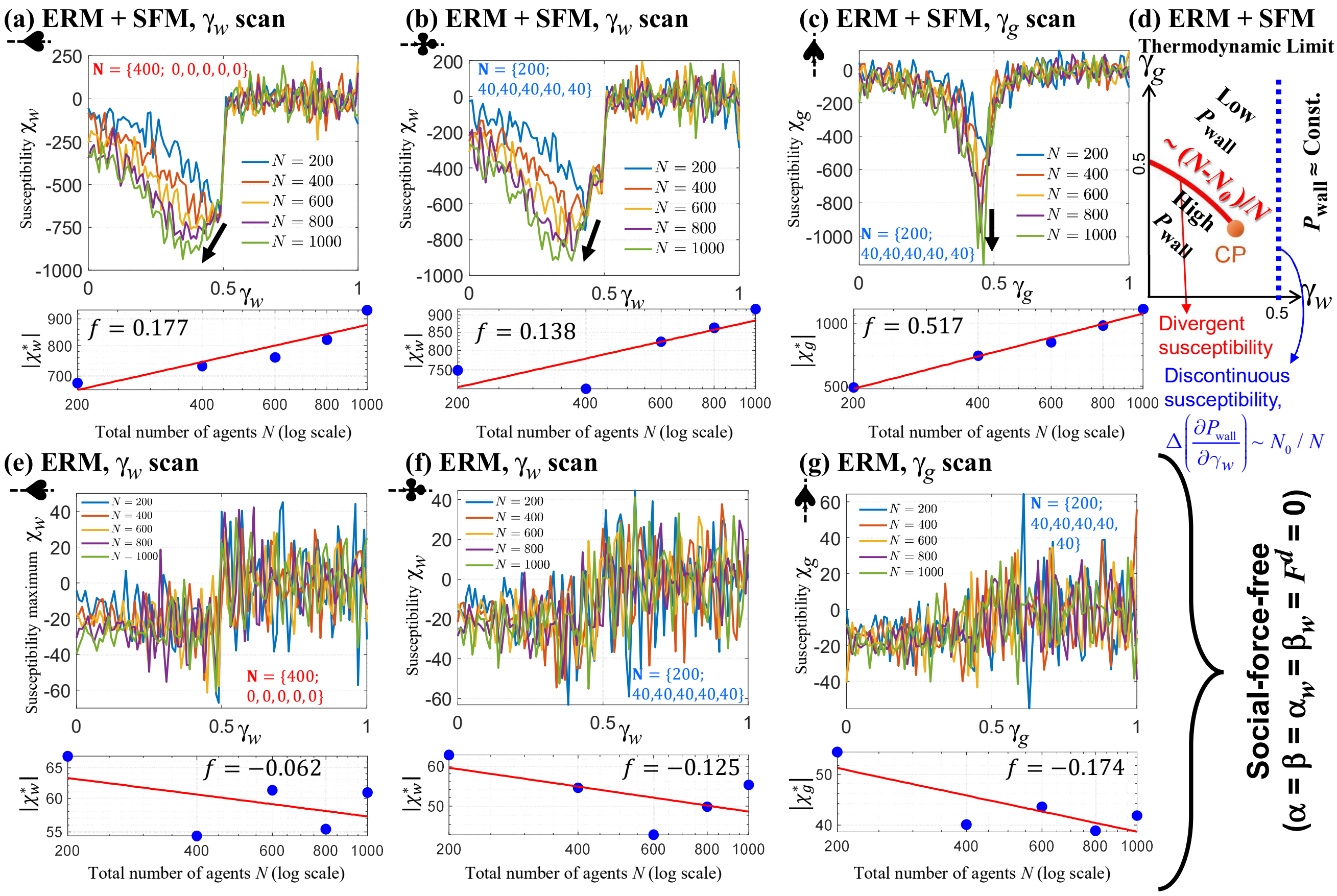

FIG. 8. (a)–(c) Susceptibilities $\chi_w$ and $\chi_g$ obtained in the ERM+SFM framework from the scans of $P_{\mathrm{wall}}$ indicated by the dashed lines with markers in Fig. 7. The lower subpanels show the finite-size scaling of the susceptibility minima, $|\chi_w^*|$ in (a) and (b) and $|\chi_g^*|$ in (c), with power-law fits (red lines) and exponents $f$. (d) Thermodynamic-limit phase diagram showing the phase-boundary segment on the EL, $\Gamma(\gamma_w, \gamma_g) = 0.5$, with divergent susceptibility (red curve), the terminating critical point (CP; red dot), and the susceptibility discontinuity at $\gamma_w = 0.5$ (blue dashed line). The phase boundary along the EL remains confined to a partial segment that vanishes in the all-independent limit, whereas the $\gamma_w = 0.5$ boundary vanishes in the fully grouped limit. The blue dashed line bounds the $P_{\mathrm{wall}} \approx$ const. region, whereas crossing the red curve leads to the high-$P_{\mathrm{wall}}$ region. (e)–(g) Corresponding results for (a)–(c), respectively, for the social-force-free ERM, $\alpha = \beta = \alpha_w = \beta_w = F^d = 0$.

At this intersection, a pronounced negative susceptibility dip $\chi_g^*$ is observed. As the system size increases from $N = 200$ to $N = 1000$, the dip deepens, as marked by the downward black arrow, and exhibits the fitted finite-size scaling $|\chi_g^*| \propto N^f$ with $f = 0.517 > 0$. This positive scaling exponent indicates that $|\chi_g^*|$ diverges in the thermodynamic limit, revealing a nonanalyticity in $P_{\rm wall}$ at $\gamma_g^*$, manifested by an increasingly sharp but continuous elevation upon entering the high-$P_{\rm wall}$ region. This behavior supports a continuous phase transition. For the $N_0 = 0$ configurations shown in Figs. 7(c1)–7(f1), we also perform power-law fits for the susceptibility dips $|\chi_g^*|$ along the scan $(\gamma_w, \gamma_g) = (0.1, \gamma_g)$. The resulting exponents are $f = 0.416$, 0.474, 0.273, and 0.327, respectively. Their consistently positive values confirm that the divergent susceptibility, and hence the thermodynamic phase boundary along the EL at $\gamma_w = 0.1$, persists across different grouped configurations.

Similar susceptibility dips $|\chi_w^*|$ are observed in the $\gamma_w$ scans shown in Figs. 8(a) and 8(b), with $|\chi_w^*| \propto N^{0.177}$ and $|\chi_w^*| \propto N^{0.138}$, respectively. Although the dip magnitudes increase with the system size, their positions shift toward $\gamma_w = 0$, as indicated by the black arrows; the extrapolated dip approaches the physical boundary $\gamma_w = 0$ with a finite magnitude. Therefore, at $\gamma_g = 0.1$, no susceptibility divergence occurs at any fixed $\gamma_w$ within the physical interval $\gamma_w \in [0, 1]$. Accordingly, only a finite segment of the EL constitutes a phase boundary characterized by the nonanalytic $P_{\rm wall}$, as depicted by the red curve in Fig. 8(d). This phase-boundary segment terminates at a critical point, represented by the red dot. The extent of the segment depends on the grouped-agent fraction $(N - N_0)/N$, but it remains confined to only part of the EL even in the fully grouped limit. In the all-independent limit $N = N_0$, this EL phase boundary disappears because the elevation of $P_{\rm wall}$ across the EL vanishes, as seen in Fig. 7(a1).

Finally, by deactivating the SFM interactions, $\alpha = \beta = \alpha_w = \beta_w = F^d = 0$, Figs. 8(e), 8(f), and 8(g) present the counterparts of Figs. 8(a), 8(b), and 8(c), respectively. In stark contrast, the susceptibility curves collapse into low-amplitude irregular oscillations, and all fitted exponents $f$ are negative, showing that the magnitude of the susceptibility minimum decreases with increasing system size. This behavior is consistent with self-averaging, as finite-size fluctuations are progressively suppressed. The susceptibility discontinuity near $\gamma_w = 0.5$ also becomes less pronounced. Neither a clear susceptibility discontinuity nor a divergence defining a phase boundary survives in the thermodynamic limit. Thus, the phase transitions summarized in Fig. 8(d) are nontrivial features that emerge only from the coupled ERM+SFM dynamics.

## IV. CONCLUSION

In conclusion, we have investigated the mechanical safety of a spatially confined crowd without egress by sequentially examining the ERM, the SFM, and the combined ERM+SFM dynamics. Our work advances crowd modeling in three key directions: (*i*) post-collision cognitive modulation, (*ii*) multi-group configurations mediated by group-induced reorientation, and (*iii*) the direct microscopic metric $\delta p_{\max}$ for quantifying the peak per-agent collision impulse. Mechanical safety was characterized using two complementary contact-based metrics: the macroscopic wall line pressure $P_{\rm wall}$, a time-coarse-grained measure of long-term wall loading, and $\delta p_{\max}$, a direct event-based measure of short-term individual collision risk. Because $P_{\rm wall}$ is insensitive to the initial speed distribution whereas $\delta p_{\max}$ retains an IC dependence, these metrics probe different aspects of mechanical risk and cannot be inferred from one another.

Within the ERM, social-group cohesion $\gamma_g$ and wall buffering $\gamma_w$ tend to retain agents within the bulk, thereby suppressing $P_{\rm wall}$. Their microscopic effects, however, are nonmonotonic. For large groups, intermediate cohesion produces a pronounced high-$\delta p_{\max}$ hazard window through clustered intra-group collisions. As $\gamma_g \to 1$, this hazard is mitigated because local pairing suppresses cluster growth, and cluster kinetic energy shifts from relative motion toward center-of-mass motion. Since only the relative component drives intra-group collisions, this shift lowers $\delta p_{\max}$. These results suggest a two-stage mitigation strategy: favor a standard-collision-biased response with $\gamma_g < 0.5$ to reduce $\delta p_{\max}$, and additionally adopt strong wall buffering with $\gamma_w > 0.5$ when suppression of $P_{\rm wall}$ is also required.

The SFM interactions produce a contrasting response. Agent-agent pushing and sliding, governed by $\alpha$ and $\beta$, significantly increase $\delta p_{\max}$, while their effects on $P_{\rm wall}$ remain comparatively modest. The coexistence of agent-wall pushing and sliding, governed by $\alpha_w$ and $\beta_w$, further magnifies $\delta p_{\max}$. Meanwhile, the active driving force enhances $P_{\rm wall}$ through near-wall accumulation. When the ERM and SFM mechanisms coexist, their competing tendencies to retain kinetic energy in the bulk or redistribute it toward the walls give rise to the wall-pressure/contact-load trade-off ($P$-$p$ trade-off). Consequently, reducing $P_{\rm wall}$ can increase $\delta p_{\max}$, and vice versa. A mechanically meaningful safety assessment therefore requires the simultaneous evaluation of both macroscopic wall loading and microscopic collision risk. The combined ERM+SFM dynamics also produces nontrivial thermodynamic phase behavior in $P_{\rm wall}$. Independent agents contribute to the phase boundary at $\gamma_w = 0.5$, where finite-size scaling reveals a susceptibility discontinuity. This boundary weakens as the independent-agent fraction decreases and vanishes in the fully grouped limit. Grouped agents, by contrast, contribute to a distinct continuous phase boundary along a finite segment of the EL $\Gamma(\gamma_w, \gamma_g) = 0.5$. Along this segment, the susceptibility diverges in the thermodynamic limit, and the phase boundary terminates at a critical point. Both phase structures disappear in the social-force-free ERM, confirming that they are emergent, non-trivial features requiring the coupled ERM+SFM dynamics.

To assess the realism and practical safety relevance of the computed $\delta p_{\max}$, we estimate the per-agent contact force as $\delta p_{\max}/\delta t_c$, where $\delta t_c$ is a representative sustained-contact duration. Taking $\delta t_c \approx 1$ s, the simulated $\delta p_{\max}$ corresponds to forces ranging from several hundred newtons to approximately $10^3$ N. Reported wall line loads in crowd disasters are on the order of $10^3$ N/m [8, 48, 49], corresponding to per-agent forces of several hundred newtons for a representative body width of 0.5 m. With model parameters chosen to reflect realistic human-scale values, the agreement between the simulated force range and the estimates inferred from reported hazardous wall line loads supports the validity of $\delta p_{\max}$ as a direct mechanical-safety metric.

Building on the present post-collision framework, a natural extension is to incorporate pre-collision avoidance mechanisms, including time-to-collision responses and predictive motion-planning methods such as VO and RVO algorithms [38–40]. Combining proactive and reactive behavioral mechanisms within a unified framework would clarify how anticipatory avoidance modifies the $P$-$p$ trade-off and the associated phase behavior. Taken together, our work demonstrates how social-group cohesion $\gamma_g$, group configuration $\mathbf{N}$, wall buffering $\gamma_w$, and the SFM interactions jointly shape $P_{\text{wall}}$ and $\delta p_{\max}$. These mechanistic insights provide actionable principles for group-configuration assignment via $\mathbf{N}$, behavioral guidance via $\gamma_g$ and $\gamma_w$, and event safety planning based on the simultaneous assessment of macroscopic $P_{\text{wall}}$ and microscopic $\delta p_{\max}$ in high-density venues without egress.

## ACKNOWLEDGMENTS

Bo-Shiun Shen is supported by the National Science and Technology Council of Taiwan under Grant. No. 114-2813-C-845-001-M.

[1] S. Ramaswamy, Annu. Rev. Condens. Matter Phys. **1**, 323 (2010).
[2] M. C. Marchetti, J. F. Joanny, S. Ramaswamy, T. B. Liverpool, J. Prost, M. Rao, and R. A. Simha, Rev. Mod. Phys. **85**, 1143 (2013), URL `https://link.aps.org/doi/10.1103/RevModPhys.85.1143`.
[3] C. Bechinger, R. Di Leonardo, H. Löwen, C. Reichhardt, G. Volpe, and G. Volpe, Reviews of modern physics **88**, 045006 (2016).
[4] G. Yu, R. Li, F. Li, J. Zhang, X. Li, Z. Chen, J. Mecke, and Y. Gao, Chinese Physics B **34**, 094702 (2025).
[5] A. P. Solon, Y. Fily, A. Baskaran, M. E. Cates, Y. Kafri, M. Kardar, and J. Tailleur, Nature physics **11**, 673 (2015).
[6] D. Helbing and P. Molnár, Physical review E **51**, 4282 (1995).
[7] D. Helbing, A. Johansson, and H. Z. Al-Abideen, Physical Review E **75**, 046109 (2007).
[8] H. Liang, L. Yang, J. Du, C.-W. Shu, and S. Wong, Transportmetrica B: Transport Dynamics **12**, 2328774 (2024).
[9] H. Liang, S. Lee, J. Sun, and S. Wong, PLoS one **19**, e0306764 (2024).
[10] T. Vicsek, A. Czirók, E. Ben-Jacob, I. Cohen, and O. Shochet, Physical review letters **75**, 1226 (1995).
[11] D. Helbing, Diplom thesis, Georg-August University, Göttingen, Germany (1990).
[12] D. Helbing, Behavioral science **36**, 298 (1991).
[13] W. Liu, J. Zhang, A. R. Rasa, and W. Song, Journal of Statistical Mechanics: Theory and Experiment **2024**, 083403 (2024).
[14] D. A. Knopoff, J. Liao, Q. Ma, and X. Yang, Mathematical Models and Methods in Applied Sciences **35**, 1637 (2025).
[15] N. Bellomo and C. Dogbe, SIAM review **53**, 409 (2011).
[16] A. Golas, R. Narain, and M. C. Lin, Phys. Rev. E **90**, 042816 (2014), URL `https://link.aps.org/doi/10.1103/PhysRevE.90.042816`.
[17] C. Burstedde, K. Klauck, A. Schadschneider, and J. Zittartz, Physica A: Statistical Mechanics and its Applications **295**, 507 (2001).
[18] M. J. Seitz and G. Köster, Physical Review E **86**, 046108 (2012).
[19] A. Seyfried, B. Steffen, W. Klingsch, and M. Boltes, Journal of Statistical Mechanics: Theory and Experiment **2005**, P10002 (2005).
[20] M. Moussaïd, D. Helbing, and G. Theraulaz, Proceedings of the National Academy of Sciences **108**, 6884 (2011).
[21] C. Von Krüchten and A. Schadschneider, Physica A: Statistical Mechanics and its Applications **475**, 129 (2017).
[22] A. Johansson, D. Helbing, H. Z. Al-Abideen, and S. Al-Bosta, Advances in Complex Systems **11**, 497 (2008).
[23] M. Moussaïd, M. Kapadia, T. Thrash, R. W. Sumner, M. Gross, D. Helbing, and C. Hölscher, Journal of the Royal Society Interface **13**, 20160414 (2016).
[24] M. Kinateder, M. Müller, M. Jost, A. Mühlberger, and P. Pauli, Applied ergonomics **45**, 1649 (2014).
[25] L. Huang, J. Gong, W. Li, T. Xu, S. Shen, J. Liang, Q. Feng, D. Zhang, and J. Sun, ISPRS International Journal of Geo-Information **7**, 79 (2018).
[26] P. A. Langston, R. Masling, and B. N. Asmar, Safety Science **44**, 395 (2006).
[27] D. Yan, G. Ding, K. Huang, C. Bai, L. He, and L. Zhang, Electronics **13**, 934 (2024).
[28] M. Moussaïd, N. Perozo, S. Garnier, D. Helbing, and G. Theraulaz, PloS one **5**, e10047 (2010).
[29] D. Helbing, I. Farkas, and T. Vicsek, Nature **407**, 487 (2000).
[30] R. Zhou, Y. Cui, Y. Wang, and J. Jiang, Tunnelling and Underground Space Technology **110**, 103837 (2021).
[31] N. Ding, Y. Zhu, X. Liu, D. Dong, and Y. Wang, Applied Mathematics and Computation **466**, 128448 (2024).
[32] Y. Gao, P. B. Luh, H. Zhang, and T. Chen, in *2013 IEEE International Conference on Automation Science and Engineering (CASE)* (IEEE, 2013), pp. 747–751.
[33] W. Wu, J. Li, W. Yi, and X. Zheng, IEEE Transactions on Intelligent Transportation Systems **23**, 15476 (2022).
[34] J. Zhang, J. Zhu, P. Dang, J. Wu, Y. Zhou, W. Li, L. Fu, Y. Guo, and J. You, International Journal of Digital Earth **16**, 1186 (2023).
[35] A. Portz and A. Seyfried, in *Pedestrian and Evacuation Dynamics* (Springer, 2011), pp. 577–586.
[36] J. Zhang, W. Klingsch, A. Schadschneider, and A. Seyfried, Journal of Statistical Mechanics: Theory and Experiment **2012**, P02002 (2012).

[37] U. Chattaraj, A. Seyfried, and P. Chakroborty, Advances in complex systems **12**, 393 (2009).
[38] P. Fiorini and Z. Shiller, The international journal of robotics research **17**, 760 (1998).
[39] J. Van den Berg, M. Lin, and D. Manocha, in *2008 IEEE international conference on robotics and automation* (Ieee, 2008), pp. 1928–1935.
[40] I. Karamouzas, P. Heil, P. Van Beek, and M. H. Overmars, in *International workshop on motion in games* (Springer, 2009), pp. 41–52.
[41] H. Liu, B. Xu, D. Lu, and G. Zhang, Applied Soft Computing **68**, 360 (2018).
[42] Z. Kang, L. Zhang, and K. Li, Applied Mathematics and Computation **348**, 355 (2019).
[43] J. Makmul, Safety Science **133**, 105006 (2021).
[44] X. Yang, R. Zhang, F. Pan, Y. Yang, Y. Li, and X. Yang, Physica A: Statistical Mechanics and its Applications **594**, 127033 (2022).
[45] Y.-C. Liu, A. Jafari, J. K. Shim, and D. A. Paley, IEEE transactions on intelligent transportation systems **23**, 16448 (2022).
[46] Q. Ji, F. Wang, and T. Zhu, in *2016 International Conference on Virtual Reality and Visualization (ICVRV)* (IEEE, 2016), pp. 317–324.
[47] D. Oberhagemann, R. Könnecke, and V. Schneider, in *Pedestrian and evacuation dynamics 2012* (Springer, 2013), pp. 1251–1258.
[48] J. Dickie and G. Wanless, Structural Engineer **71**, 216 (1993).
[49] R. Smith and L. Lim, Safety science **18**, 329 (1995).